\documentclass[12pt]{iopart}

\usepackage{graphicx}
\usepackage{epsfig}

\usepackage{amssymb}

\begin{document}



\title[Ab-initio calculation of electronic stopping power along$\dots$]{Ab-initio calculation of electronic stopping power along glancing swift heavy ion tracks in perovskites}


\author{O.~Osmani$^{1}$,~A.~Duvenbeck$^{1}$,~E.Akc\"oltekin$^{1}$,~R.~Meyer$^{1}$,~H.~Lebius$^{2}$ and M.~Schleberger$^{1}$}
\address{$^{1}$~Department of Physics, University of Duisburg-Essen, D-47048 Duisburg, Germany}
\address{$^{2}$~CIMAP, blvd Henri Becquerel, 14070 Caen, France}
\ead{marika.schleberger@uni-due.de}

\begin{abstract}
In recent experiments the irradiation of insulators of perovskite type with swift heavy ions under glancing incidence has been shown to provide a unique means to generate periodically arranged nanodots at the surface. The physical origin of these patterns has been suggested to stem from a highly anisotropic electron density distribution within the bulk. In order to show the relevance of the electron density distribution of the target we present a model calculation for the system Xe$^{+23}$ $\rightarrow$~SrTiO$_{3}$ that is known to produce the aforementioned surface modifications. On the basis of the Lindhard model of electronic stopping, we employ highly-resolved \emph{ab-initio} electron density data to describe the conversion of kinetic energy into excitation energy along the ion track. The primary particle dynamics are obtained via integration of the Newtonian equations of motion that are governed by a space- and time-dependent friction force originating from Lindhard stopping. The analysis of the local electronic stopping power along the ion track reveals a pronounced periodic structure. The periodicity length strongly varies with the particular choice of the polar angle of incidence and is directly correlated to the experimentally observed formation of periodic nanodots at insulator surfaces. 
\end{abstract}
\pacs{61.80.Jh,61.80.Az,61.82.Ms}
\vspace{2pc}
\noindent{\it Keywords}:~electronic stopping, swift heavy ions, incidence, perovskite, insulator, thermal spike
\vspace{2pc}
\submitto{\JPCM}
\maketitle

\section{Introduction}
\label{sec:introduction}
The irradiation of a solid surface with a $\sim$100~MeV ion beam leads to the formation of cylindrical tracks within the target material (see e.g. \cite{Itoh2001,Toulemonde2006a,Weidinger2004}). This phenomenon can be explained in terms of kinetic excitation of electrons along the path of the penetrating ion. These electronic excitations are directly connected with structural changes within the damage zone which may also extend to the very surface, for instance, manifesting themselves in the formation of nanodots,  for instance.\\
In the case of normal incidence of the ion beam the nanodots are statistically distributed over the entire irradiated area \cite{Khalfaoui2005,Garcia-Navarro2006,Mueller2003,Skuratov2003} due to the random distribution of the ion impact sites. In recent experiments~\cite{Akcoltekin2007}, the irradiation of  perovskite crystals (ABO$_{3}$) under \textit{glancing incidence} has been shown to provide a unique means to generate chain-like, periodic nanodots at the surface, the extension of which varies in the range of a few hundred nanometers up to some micrometers depending on the particular angle of incidence. This finding has been ascribed to the specific anisotropic electron density of the  perovskite  structure. The nanodots can be imaged by atomic force microscopy \cite{Khalfaoui2006}, for instance, and may be regarded as a fingerprint of the electron density projected onto the surface.\\
Although the detailed microscopic mechanisms of this kind of defect formation are not entirely understood, the anisotropy of the electron density inducing a discontinuous kinetic energy dissipation of the ion appears to play a key role. For many practical purposes the energy loss per track length can be calculated using SRIM \cite{Srim2003web,Biersack_80} or equivalent standard codes. However, these codes are not capable to account for the detailed electronic structure and are limited to perpendicular incidence. In addition, these codes do not provide any space-resolved information on energy losses inside the crystal, which are supposed to be essential in our case.\\
Therefore this paper aims at a detailed investigation of the spatial electronic energy loss under $\sim$100~MeV glancing incidence irradiation. For the model system Xe$^{+23}$~$\rightarrow$~SrTiO$_{3}$ which has been shown to produce the aforementioned periodic nano patterns, we combine \emph{ab-initio} electron density computations with trajectory calculations in order to study the electronic stopping dynamics along the ion track for different polar and azimuthal angles of incidence. 
\section{Model}
\label{sec:model}
\begin{figure}
	\centering
		\includegraphics[width=10.5cm, height=10.5cm]{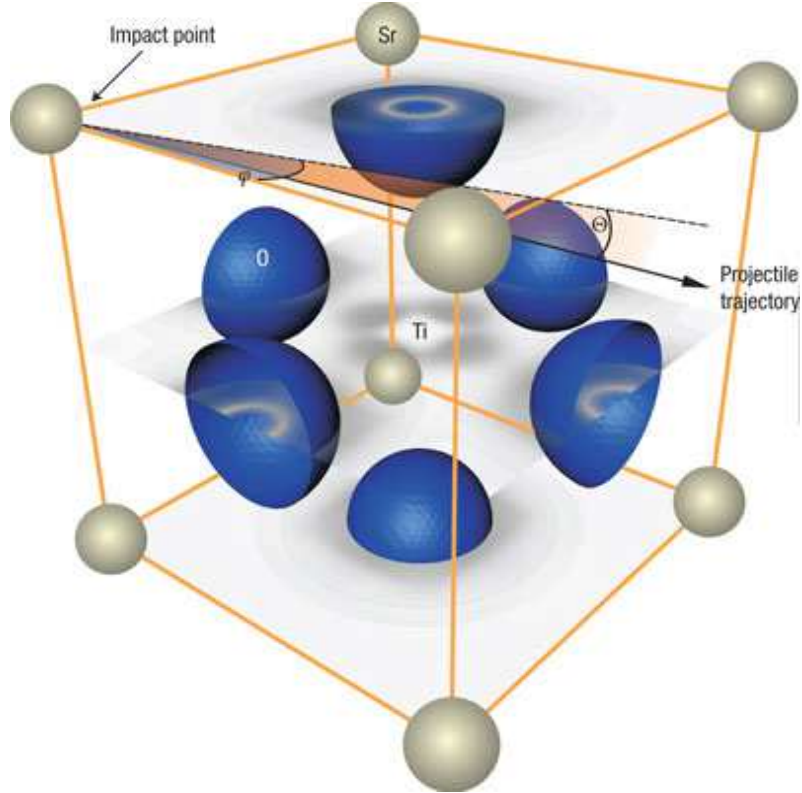}
		\caption{[figure and caption reprinted from Ref.~\cite{Akcoltekin2007} ] DFT calculation of the electron density of SrTiO$_{3}$: The grey shading represents electron density. Atom positions (apart from the central titanium atom) are also visualized. The arrow indicates a possible projectile trajectory defined by the azimuthal angle $\varphi$ as well as the incidence angle $\Theta$.}
	\label{fig:figure1}
\end{figure}
For the irradiation of SrTiO$_{3}$ (100) with Xe$^{+23}$ ions of $\sim$100~MeV of kinetic energy electronic stopping dominates over nuclear stopping by two orders of magnitude. In this energy regime electronic stopping occurs via direct inelastic projectile-electron collisions and is usually treated in terms of the Lindhard model \cite{Lindhard_61} yielding an electronic stopping cross section  $S_{e}=\beta \cdot v_{p}$ with a material parameter $\beta$ depending on the specific target-projectile combination and $v_{p}$ denoting the projectile velocity. For a target containing different elements $i$ with individual $\beta^{(i)}$'s the effective electronic stopping cross section $S_{e,\mbox{\tiny {eff}}}$ is given by
\begin{equation}
S_{e,\mbox{\tiny {eff}}}=\beta_{\mbox{\tiny {eff}}}\cdot v_{p}=\sum_{i}w^{(i)}\beta^{(i)}\cdot v_{p}
\end{equation}
with relative weight factors $w^{(i)}$ mirroring the stoichiometry of the target. The electronic energy loss per track length (stopping power) of the projectile along its path through the solid then reads
\begin{equation}
\frac{dE}{dx}\big |_{e}=N_{e}(\vec{r}_{p}(t))\cdot \beta_{\mbox{\tiny {eff}}}\cdot |\dot{\vec{r}}_{p}(t)| \mbox{~,}
\label{eq:equation1}
\end{equation}
where $N_{e}(\vec{r}_{p}(t))$ denotes the electron density of the target at the time-dependent position $\vec{r}_{p}(t)$ of the penetrating ion. It should be mentioned that for the system studied here the initial charge of $q=+23$ is close to the equilibrium charge state during the passage of the ion through the target \cite{Shima1983,Shima1989}.
\begin{figure}
	\centering
		\includegraphics[width=10.5cm, height=10.5cm]{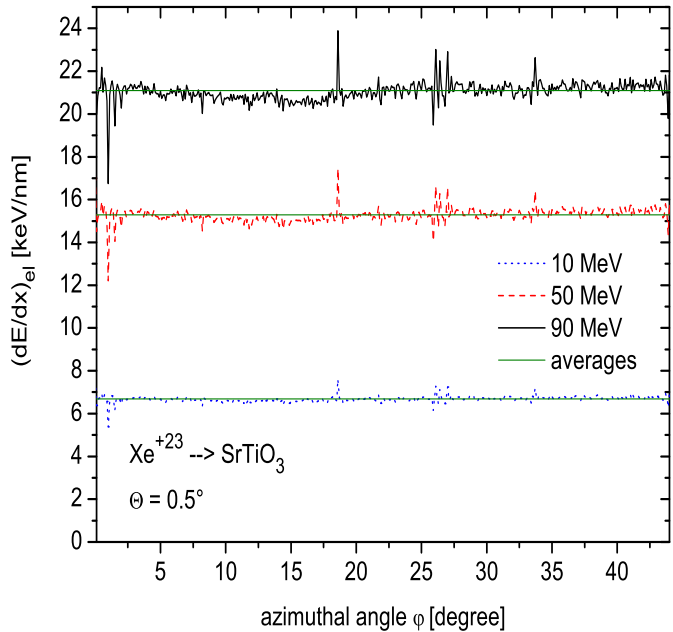}
		\caption{Average electronic stopping power $(dE/dx)_{el}$ for different irradiation energies (10~MeV,~50~MeV,~90~MeV) and a polar angle of $0.5^{~\circ}$. The step width $\Delta \varphi$ has been set to 0.1~$^{\circ}$.}
	\label{fig:figure2}
\end{figure}\\
The electron density $N_{e}(\vec{r})$ constituting the \textit{first} essential input parameter in Eq.~(\ref{eq:equation1}) is calculated for one unit cell of the perovskite using the \textsc{Abinit} package~\cite{Gonze2002} with pseudopotentials from the Fritz-Haber-Institut \cite{Fuchs1999}. Employing a cutoff energy of 96~Hartree with a sampling of 512~$k$-points on a cubic grid, the calculations have been carried out within the GGA approximation using the PBE functional \cite{Perdew1996}. The lattice constant of SrTiO$_{3}$ has been determined as 3.905~\AA~by minimization of the energy. It should be noted that ab-initio calculations using pseudopotentials  only give the electron density of the valence electrons. The core electrons, however, are strongly localized around the nuclei which limits their influence to a very small fraction of the volume. In the energy range we are considering here, the projectile comes close to the nuclei only very rarely. Therefore consideration of the core electrons would lead to minor quantitative changes of our result, only, and no qualitative changes at all.\\

The electron density $N_{e}(\vec{r})$ resulting from this \emph{ab-initio} computation is shown in Fig.~\ref{fig:figure1}. The majority of the electrons are localized around the oxygen atoms. Moreover, the electron density in the TiO$_{2}$ plane clearly exceeds the one observed in the SrO plane. Thus, it becomes directly obvious that $N_{e}(\vec{r})$ varies on a sub-\AA ngstr\"om scale and hence should not be treated as a homogeneous electron gas.\\
\begin{figure}
	\centering
		\includegraphics[width=10.5cm, height=10.5cm]{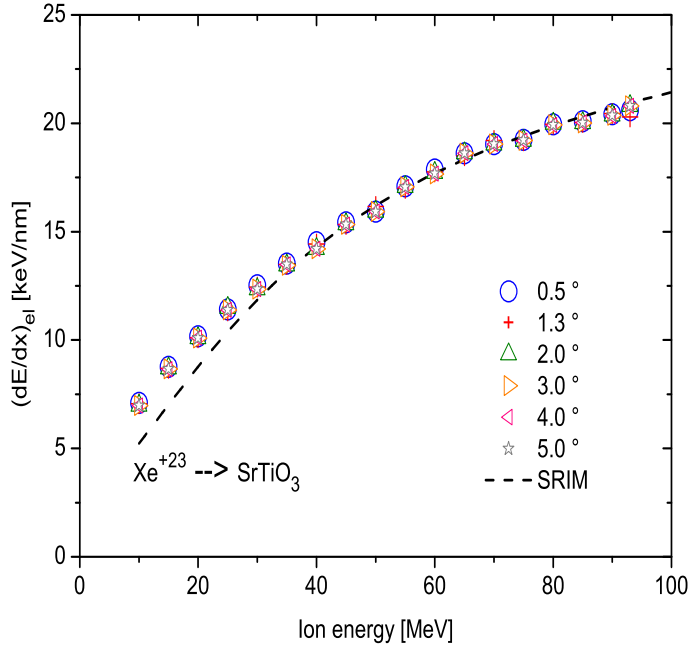}
		\caption{Stopping power as a function of ion energy for different polar angles of incidence. For each ion energy the stopping power has been averaged over the entire range of the azimuthal angle. The dashed line represents the calculated energy loss using SRIM~\cite{Srim2003web}.}
	\label{fig:figure3}
\end{figure}
The \emph{second} parameter to be inserted in Eq.~(\ref{eq:equation1}) is the trajectory $\vec{r}_{p}(t)$ and its first time derivative. The trajectory is calculated by numerical integration of the Newtonian equations of motion for the projectile. In view of the fact that energy dissipation is governed by electronic stopping, in a first-order approximation no interatomic potentials have to be incorporated into the equations of motion. Thus, the only interaction between the primary ion and the target is an effective friction force according to Eq.~(\ref{eq:equation1}) that originates from Lindhard stopping. During the time integration of the ion track, the energy loss per numerical time step is output for further analysis.
\section{Results}
\begin{figure*}
	\centering	
		\includegraphics[width=14cm, height=14cm]{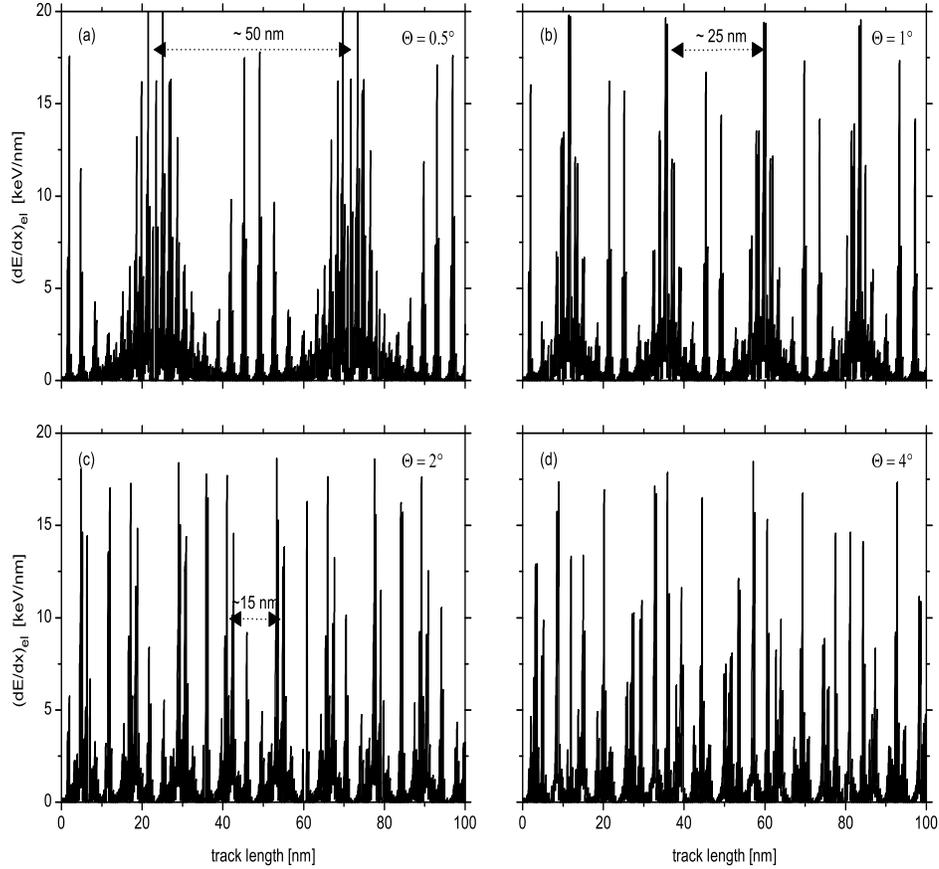}
		\caption{Electronic stopping power as a function of track length, i.e. travelled distance of the ion from the point of impact. Data has been calculated for a Xe$^{+23}$ ion of 93~MeV initial kinetic energy penetrating a SrTiO$_{3}$ target under polar angles $\Theta=0.5^{\circ}$~(a), $\Theta=1^{\circ}$~(b), $\Theta=2^{\circ}$~(c) and $\Theta=4^{\circ}$~(d), respectively. The arrows represent the periodicity lengths of subsequent peaks of same type as explained in detail in the text.}
	\label{fig:figure4}
	\end{figure*}

The computational procedure as outlined in the previous section has been performed for the irradiation of SrTiO$_{3}$ with Xe$^{+23}$ ions in the energy range from (10-100)~MeV. For each bombarding energy the azimuthal angle of impact has been varied. The crystal symmetry of the target allows to limit the azimuthal variation from 0~to~45 degrees. The impact point is chosen as illustrated in Fig.~\ref{fig:figure1} and kept constant for all calculations presented here. However, preliminary computations have shown a negligible influence of the choice of impact point particularly with regard to the results presented here.\\
Figure~\ref{fig:figure2} shows the dependence of the electronic stopping power averaged along the trajectory on the azimuthal angle of incidence for exemplary irradiation energies of 10, 50 and 90~MeV. The polar angle has been chosen as $\Theta=0.5^{~\circ}$ with respect to the surface plane as indicated in Fig.~\ref{fig:figure1}. Considering the obtained stopping powers we observe rather moderate fluctuations ($\sigma \approx 3\%$) around the mean values of 21.1~keV/nm (90~MeV), 15.3~keV/nm (50~MeV) and 6.7~keV/nm (10~MeV), respectively. The obtained finestructures in the angular dependence of the stopping power are directly related to the inhomogeneous electron density distribution and exhibit identical curve characteristics independent of the ion energy. The pronounced peaks at 19$^{\circ}$, 27$^{\circ}$ and 34$^{\circ}$ correspond to low-indexed crystallographic directions for the chosen impact point. Close to the [001] direction minima in the electronic stopping are due to the presence of the Sr atom. The ratio of the amplitudes between the different projectile energies scales with $\sqrt{E}$ due to the employed Lindhard stopping cross section. \\
In order to verify the accuracy of our approach we calculate the stopping power averaged over the azimuthal degree of freedom for different ion energies and compare these values with SRIM calculations. The results are depicted in Fig.~\ref{fig:figure3} for various polar angles of incidence $\Theta$. It becomes directly obvious that the stopping power is largely independent of the angular degrees of freedom. In the energy range from (30-80)~MeV the computed stopping powers are in accordance with SRIM data. For impact energies below 30~MeV, however, we observe a systematic overestimation compared to the SRIM results. This discrepancy clearly stems from the fact that in this energy regime the electronic stopping power is not acurrately described by a $\sqrt{E}$-dependence as assumed in our case.\\
Summarizing up to here, the averaged electronic stopping power appears to be independent of the particular choice of angular impact parameters. However, experiments cleary reveal the explicit dependence of the geometry of nano surface dots (e.g. total chain length, periodicity length) on the particular choice of polar angle. Therefore, it is necessary to take a space-resolved look on the energy transfer characteristics along the ion track for different polar angles.\\
 Figure~\ref{fig:figure4} illustrates the electronic stopping power along the 93~MeV Xe$^{23+}$ ion track in SrTiO$_{3}$ for $\Theta=0.5^{\circ}$, $\Theta=1^{\circ}$, $\Theta=2^{\circ}$ and $\Theta=4^{\circ}$, respectively. For the purpose of clarity the depicted trajectory length is limited to 100~nm. The azimuthal angle is chosen as $\varphi=7^{\circ}$ representing a highly indexed direction. For $\Theta=0.5^{\circ}$ (see Fig.~\ref{fig:figure4}~(a)) we notice two pronounced peaks at 20~nm and 70~nm track length. In the middle of those peaks we detect more thinned-out secondary peaks of approximately the same half width contributing about 35~\% to the total excitation energy. Each of the peaks exhibits a finestructure of $\delta$-shaped peaks naturally resulting from fluctuating electron densities along the track.\\
  At this point we remark that a higher resolved representation of the data would mirror the energy deposition during the passage of each unit cell. However, in our case the main point of interest is the observed distance between two consecutive areas of increased electronic stopping. The periodicity length derived for $\Theta=0.5^{\circ}$ is approximately 50~nm. For a slightly increased polar angle of 1$^{\circ}$ (see Fig.~\ref{fig:figure4}~(b)) the same kind of peaks can be identified, however, each kind with a reduced periodicty of approximately 25~nm. For even larger polar angles of $\Theta=2^{\circ}$ and $\Theta=4^{\circ}$, respectively, the two types of peaks begin to merge leading to a nearly constant enveloping function. Thus, the pronounced peak structure as seen before is increasingly smeared out and the periodicity tends to disappear.\\
These results qualitatively explain the experimental findings that the periodic nanodot formation is limited to glancing angles of incidence up to a few degree.\\
We are well aware that a more quantitative treatment of defect formation using our model must incorporate a realistic description of the transport of the excitation energy as well as the coupling of the electronic system to the lattice. This may be achieved in terms of a two temperature approach \cite{Anisimov66} as employed in the thermal spike model \cite{Toulemonde2006a}, for instance. However, in our case the anisotropy of the electron density and the necessity to properly implement the surface leads to a break-down of the radial symmetry usually presumed in standard thermal spike calculations and, thus, to a drastical increase in computational effort to numerically solve the corresponding transport equations. Calculations of that type are currently under way in our lab.
\section{Conclusion}
We have presented a model that uses highly-resolved \emph{ab-initio} electron density data in combination with molecular dynamics in order to obtain space-resolved electronic excitation characteristics along the track of swift heavy ions in matter. The model has been applied to the irradiation of SrTiO$_{3}$ with swift Xe$^{+23}$ ions in glancing incidence geometry for different ion energies and angular configurations of incidence.\\
Results for the average electronic stopping power are in good agreement with SRIM and, moreover, do not show a significant influence of the particular choice of the angular degree of freedom. However, the specific choice of the polar angle turns out to be crucial for the spatial distribution of electronic excitation along the ion track. A detailed analysis of the local behaviour of the electronic stopping power along the trajectory of the primary particle reveals a pronounced periodic structure that qualitatively explains the physical origin of periodic nanodot formation at insulator surfaces.\\
Finally, we would like to point out that our approach is applicable to any insulating material for which the electron density data can be calculated.\\
\begin{center}
\textbf{Acknowledgment}
\end{center}The authors would like to acknowledge financial support from the Deutsche Forschungsgemeinschaft within the Sonderforschungsbereiche 616 entitled \emph{Energy Dissipation at Surfaces} and 445 entitled \emph{Nano Particles from the Gas Phase}.
\section*{References}
\bibliographystyle{unsrt}
\bibliography{Literaturdatenbank}
\end{document}